\documentclass[aip, graphicx, reprint]{revtex4-1}

\usepackage{graphicx}
\usepackage[utf8]{inputenc}
\usepackage[T1]{fontenc}

\newcommand{\degree}{$^{\circ}$}
\newcommand{\unit}[1]{\ \text{#1}}

\usepackage{color}

\begin{document}


\title{Micromagnetic modelling of magnetic domain walls in curved cylindrical nanotubes and nanowires}



\author{L. Skoric}
\email{ls604@cam.ac.uk}
\affiliation{Cavendish Laboratory, University of Cambridge, UK}

\author{C. Donnelly}
\affiliation{Cavendish Laboratory, University of Cambridge, UK}

\author{C. Abert}
\affiliation{Faculty of Physics, University of Vienna, Vienna, Austria}
\affiliation{Research Platform MMM Mathematics-Magnetism-Materials, University of Vienna, Austria}

\author{A. Hierro-Rodriguez}
\affiliation{SUPA, School of Physics and Astronomy, University of Glasgow, UK}
\affiliation{Depto. Física, Universidad de Oviedo, 33007 Oviedo, Spain}

\author{D. Suess}
\affiliation{Faculty of Physics, University of Vienna, Vienna, Austria}
\affiliation{Research Platform MMM Mathematics-Magnetism-Materials, University of Vienna, Austria}

\author{A. Fern\'andez-Pacheco}
\email{amalio.fernandez-pacheco@glasgow.ac.uk}
\affiliation{Cavendish Laboratory, University of Cambridge, UK}
\affiliation{SUPA, School of Physics and Astronomy, University of Glasgow, UK}

\date{\today}

\begin{abstract}
    We investigate the effect of curvature on the energy and stability of domain wall configurations in curved cylindrical nanotubes and nanowires. We use micromagnetic simulations to calculate the phase diagram for the transverse wall (TW) and vortex wall (VW) states in tubes, finding the ground state configuration and the metastability region where both types of walls can exist.
    The introduction of curvature shifts the range for which the TW is the ground state domain wall to higher diameters, and increases the range of metastability. We interpret this behavior to be primarily due to the curvature-induced effective Dzyaloshinskii–Moriya term in the exchange energy.
    Furthermore, we demonstrate qualitatively the same behavior in solid cylindrical nanowires.
    Comparing both tubes and wires, we observe how while in tubes curvature tends to suppress the transformation from the TW to VW, in wires it promotes the transformation of the VW containing the Bloch point into the TW. These findings have important implications in the fundamental understanding of domain walls in 3D geometries, and the design of future domain wall devices.
\end{abstract}


\maketitle 

Magnetic domain walls (DWs) in nanostructures are topological solitons that have sparked the curiosity of researchers since the early days of micromagnetism.
Initially, studies focused on thin films of magnetic material, referred to as strips. The ease of fabrication and characterization made these systems an ideal platform for classifying DW structures \cite{mcmichaelHeadHeadDomain1997, nakataniHeadtoheadDomainWalls2005, rougemaillePhaseDiagramMagnetic2012}, and investigating their dynamics under fields \cite{beachDynamicsFielddrivenDomainwall2005,shigetoInjectionMagneticDomain1999, cowburnDomainWallInjection2002,obrienDynamicPropagationNucleation2011}, spin currents\cite{thiavilleMicromagneticUnderstandingCurrentdriven2005,emoriCurrentdrivenDynamicsChiral2013,beachCurrentinducedDomainWall2008}, spin waves \cite{hanMagneticDomainwallMotion2009, pirroExperimentalObservationInteraction2015,kimInteractionPropagatingSpin2012,parkInteractionSpinWaves2020} and thermal effects \cite{islamThermalGradientDriven2019, leliaertThermalEffectsTransverse2015}.
The continuous research in these systems has been warranted by a number of potential applications such as memory and logic \cite{allwoodMagneticDomainWallLogic2005,parkinMagneticDomainWallRacetrack2008, luoCurrentdrivenMagneticDomainwall2020}, sensing \cite{mattheisConceptsStepsRealization2012,borieGeometricalDependenceDomainWall2017}, and thermomagnetic devices \cite{bauerSpinCaloritronics2012,krzysteczkoDomainWallMagnetoSeebeck2015}.

There has been a recent push in nanomagnetism towards extending the quasi-2D geometry of strips to three-dimensional (3D) structures.
3D nanomagnetism introduces new physics, controlling and enhancing the magnetic properties by changing the system's geometry \cite{fernandez-pachecoThreedimensionalNanomagnetism2017, streubelMagnetismCurvedGeometries2016}.
In this regard, cylindrical nanowires (NWs) and nanotubes (NTs) have proven to be fascinating systems \cite{stanoChapterMagneticNanowires2018}.
Due to the change in topology, DWs in these systems can have properties distinct from their 2D counterparts. They can be massless, moving without intrinsic pinning and remaining stable even at speeds exceeding the Walker limit \cite{yanBeatingWalkerLimit2010,yanFastDomainWall2011,hertelUltrafastDomainWall2016, schobitzFastDomainWall2019,maCherenkovtypeThreedimensionalBreakdown2020}.

The advancement of nanofabrication and computational methods make studying increasingly more complex geometries possible \cite{skoricLayerbyLayerGrowthComplexShaped2020}.  An important geometrical quantity is curvature, which has been shown to have a strong effect on the magnetic energy landscape, producing curvature-induced anisotropy, and chirality via an effective Dzyaloshinskii–Moriya interaction (DMI) \cite{streubelMagnetismCurvedGeometries2016, hertelCurvatureinducedMagnetochirality2013,shekaNonlocalChiralSymmetry2020}. In cylindrical nanowires, the introduction of curvature has been shown to, among others, lead to curvature-induced chiral pinning potentials \cite{yershovCurvatureinducedDomainWall2015}, and controllable DW oscillation \cite{morenoOscillatoryBehaviorDomain2017a, cacilhasControllingDomainWall2020}.

With promising prospects for magnetic domain walls in three-dimensional systems for novel and enhanced devices, as well as interesting physical properties, obtaining a detailed understanding of the influence of curvature on the domain walls is key. Here we determine the effect of curvature on domain walls in nanotube and nanowire structures. We reveal that the introduction of curvature can change the stability of domain walls and their preferred configuration, leading to spontaneous topological transformations in a 3D conduit.

There are many common features in the configuration and dynamics of domain walls in NWs and NTs, despite them being topologically distinct structures. In NTs, there are two stable DW configurations: the transverse wall (TW) with a component of the magnetization transverse to the axial direction\cite{nielschSwitchingBehaviorSingle2002,forsterDomainWallMotion2002}, and the vortex wall (VW) with azimuthal magnetization\cite{landerosReversalModesMagnetic2007}. While both walls can exist for a wide range of cylinder diameters, the TW is favorable in narrow geometries and the VW in wide ones\cite{fergusonMetastableMagneticDomain2015}.

\begin{figure*}[ht!]
    \centering
    \includegraphics[width=\linewidth]{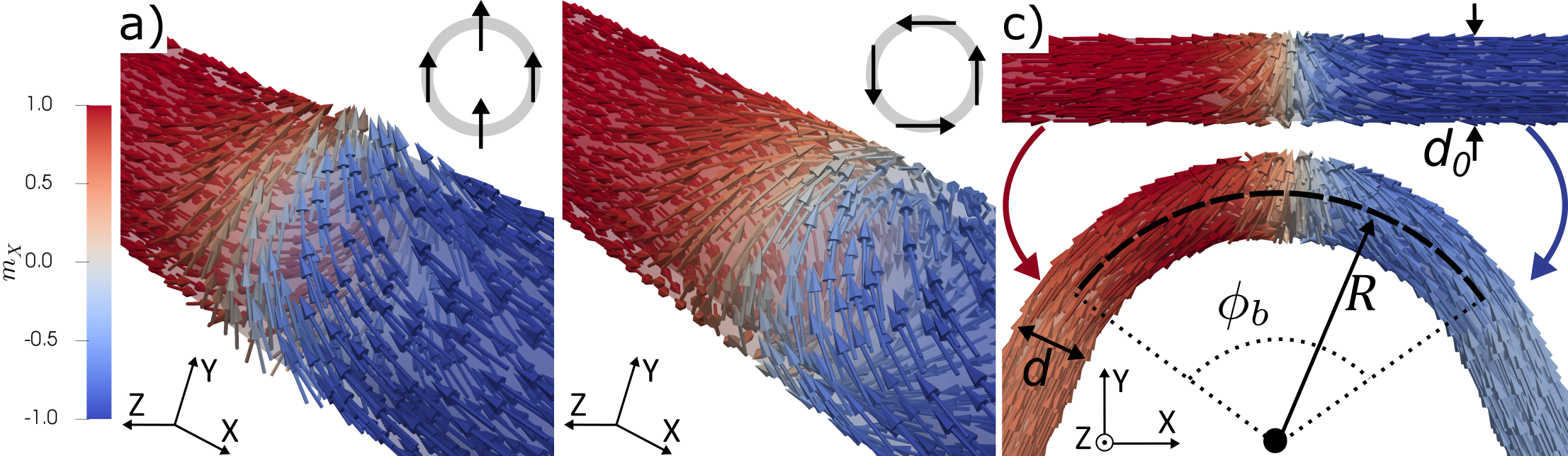}
    \caption{Domain wall initialization in magnetic nanotubes.
        (a) A transverse wall (TW) and (b) vortex wall (VW) are initialized in straight tubes. Colours indicate the magnetization component along the tube axis. Insets show the cross-sectional schematic of the tube in the centre of the DW.
        c) The mesh of the straight tube with diameter $d_0$ is bent by an angle $\phi_b = 130$\degree and squeezed to get a tube with radius of curvature $R_C$ and diameter $d$ while preserving the total length. The magnetization is smoothly transferred to the curved geometry.    }
    \label{fig:init}
\end{figure*}

The DWs in NWs share the same external spin texture as those in NTs, but develop unique features due to an additional degree of freedom. At the center of the wire, the VW leads to the formation of a Bloch point (BP), a singularity where magnetization locally cancels out \cite{dacolObservationBlochpointDomain2014, wartelleBlochpointmediatedTopologicalTransformations2019}. Additionally, the TW in thick tubes and wires with a larger diameter develops additional curling structure and is often referred to as a transverse-vortex wall (TVW) \cite{stanoChapterMagneticNanowires2018, jametHeadtoheadDomainWalls2015}. The TVW and TW share the same topology, as the curling can develop by a continuous transformation of the transverse feature \cite{stanoChapterMagneticNanowires2018}.
For the sake of easier comparison between NTs and NWs, in this article we refer to the Bloch point wall (BPW) and TVW as VW and TW respectively, addressing the differences where necessary.

We investigate the effect of curvature on the energy and stability of DWs by performing zero-temperature finite-element micromagnetic simulations using the magnum.fe library \cite{abertMagnumFeMicromagnetic2013}. We simulate the structures as permalloy with $M_S = 8\times 10^5 \unit{Am}^{-1}$, $A = 1.3 \times 10^{-11} \unit{Jm}^{-1}$ and no magneto-crystalline anisotropy.
We study a range of diameters and curvatures for both NTs and NWs in order to map the phase diagram of the systems and determine the influence of curvature on the domain wall properties.

We first consider the case of hollow magnetic NTs. Either TW or VW (Fig. \ref{fig:init}a,b) are initialized in the center of a straight NT with outer diameter $d_0=30$ nm, inner diameter $d^{in}_0=0.8 \times d_0=24 \unit{nm}$, and length $L=800 \unit{nm}$. The states are relaxed by integrating the Landau-Lifshitz-Gilbert equation (LLG) with high damping $\alpha = 1$ for faster convergence.

To systematically initialize metastable states in different geometries, we perform coordinate transformations on the initial mesh and the corresponding transformations on the magnetization (see Supporting Information S1).
The initial model is meshed with GMSH \cite{geuzaineGmsh3DFinite2009} with the maximum mesh element size set to 2.1 nm. This is significantly lower than the permalloy dipolar exchange length of $\ell = \sqrt{2A/\mu_0M_S^2} = 5.7 \unit{nm}$, and ensures that the node spacing stays below $\ell $ after the transformations (see Supporting Information S2).
We let the magnetization relax in the new geometry before calculating the energy of the final state.
In particular, we map a range of NT diameters $d=14-50\unit{nm}$ by scaling the structure uniformly in the radial direction while keeping the relative thickness constant ($d^{in}/d=0.8$). We further introduce curvature by bending the tube in the middle by an angle $\phi_b=130$\degree, creating a region of constant radius of curvature $R_C$ ranging from 40 to 200 nm (Fig. \ref{fig:init}c).

\begin{figure*}[ht]
    \centering
    \includegraphics[width=\linewidth]{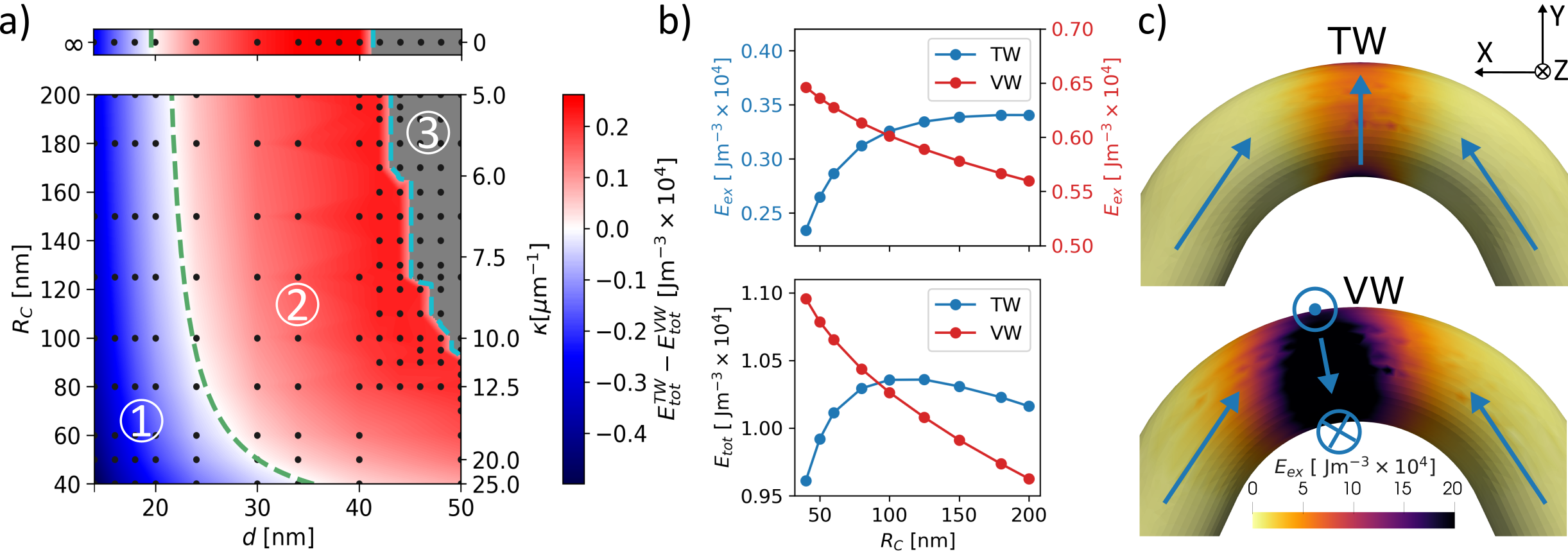}
    \caption{
    Domain wall curvature dependence in hollow nanotubes.
    (a) Phase diagram of the energy density difference between the TW and VW ($E^{TW}_{tot}-E^{VW}_{tot}$) as a function of radius of curvature $R_C$ (or curvature $\kappa$) and nanowire diameter $d$. The dashed green line marks the boundary where both DWs are of equal energy. The separated section of the diagram with $R_C = \infty$ refers to the straight wires. Points denote simulation data while the rest of the diagram is obtained by linear interpolation. In the blue region (1), TW is the lower energy domain wall, and VW is metastable ($E^{TW}_{tot}<E^{VW}_{tot}$). In the red region (2), the VW is the favorable state, and the TW is metastable ($E^{VW}_{tot}<E^{TW}_{tot}$). In the gray region (3) separated by the cyan line from region (2), the VW is the only stable state, and the initialized TW collapses to a VW.
    (b) Exchange energy density (top) and total magnetic energy density (bottom) of the TW and VW configurations as a function of $R_C$ in a nanotube with diameter $d=24\unit{nm}$. In the exchange energy plot, the TW and VW exchange energies are shifted vertically with respect to each other for easier comparison of the trends followed for the two types of walls.
    (c) Side view of the nanotube colored by the local exchange energy density for TW (top) and VW (bottom), with magnetization direction schematically shown by blue arrows. The rotation of magnetization against the curvature on one side of the VW results in an increased exchange energy when compared to the TW.}
    \label{fig:tube_diagram}
\end{figure*}

We study the relative stability of the DW types at zero temperature by plotting the diagram of the difference in total magnetic energy density between the TW and the VW ($E^{TW}_{tot}-E^{VW}_{tot}$) for a range of diameters and curvatures (Fig. \ref{fig:tube_diagram}a). Considering the energy \textit{density} instead of the total energy allows us to directly compare the wires of different diameters \cite{fergusonMetastableMagneticDomain2015}. Moreover, the \textit{difference} in energy densities will depend only on the domain wall, as the magnetization configuration is the same in other parts of the wire.

We start by considering the simple case of straight wires which we denote by the infinite radius of curvature $R_C=\infty$ (see upper section of Fig. \ref{fig:tube_diagram}a). We observe three main regions. First, for diameters $d<20\unit{nm}$ (blue region), the TW is energetically favorable ($E^{TW}_{tot}<E^{VW}_{tot}$). Second, in NTs of larger diameter, the VW is preferred, while TW exists in a metastable state (red region, $E^{VW}_{tot}<E^{TW}_{tot}$). Third, for diameters larger than 42 nm, the TW is no longer metastable and collapses to the VW which is now the only stable DW configuration (gray region).

The observed influence of the diameter on the energies of TW and VW in NTs agrees with the previous analytical and numerical studies \cite{landerosDomainWallMotion2010,landerosReversalModesMagnetic2007, fergusonMetastableMagneticDomain2015} and can be understood as a competition between the exchange and magnetostatic energies. The VW configuration minimizes the magnetostatic energy as the magnetization is tangential to the surface of the wire. On the other hand, the TW has a lower exchange energy as the spins in each cross-section are largely parallel. As the diameter decreases, the exchange term becomes dominant, favoring the TW state.

Introducing curvature into the system has a significant impact on the energy landscape, as shown in the wider phase diagram in Fig. \ref{fig:tube_diagram}a. The introduction of curvature in general acts to preferentially stabilize the TW over the VW, as can be seen by the increase in the blue region with decreasing radius of curvature. Both boundaries between the regions observed for straight wires, namely the one evaluating which wall is energetically favorable (green line) and the one where the TW is no longer metastable (cyan line), shift towards higher diameters. Notably, for a minimum considered radius of curvature $R_C=40\unit{nm}$, the critical diameter between the two DW configurations moves from 20 nm for straight wires to 35 nm, a significant change in behavior. Furthermore, for radii of curvature $R_C < 90\unit{nm}$ we find that the TW can exist in a metastable state for the full range of diameters investigated.

To understand the shift of boundaries as a function of curvature, we consider the evolution of the total energy density ($E_{tot}=E_{ex}+E_{demag}$) and the exchange energy density $E_{ex}$ for a NT with diameter $d=24 \unit{nm}$ (Fig. \ref{fig:tube_diagram}b). We first consider the exchange energy in Fig. \ref{fig:tube_diagram}b (top plot). In the case of the TW, introducing curvature reduces the wall exchange energy since the total rotation angle between the two domains is smaller (see Fig. \ref{fig:tube_diagram}c top). This is often understood as the chiral symmetry breaking due to the effective DMI terms in exchange energy in curvilinear coordinates \cite{yershovCurvatureinducedDomainWall2015, shekaNonlocalChiralSymmetry2020,shekaCurvatureEffectsStatics2015}. On the other hand, the rotational symmetry of the VW means that both the curvature preferred and unpreferred transitions are present, and the magnetization rotates against the curvature on one side of the wall (bottom panel of Fig. \ref{fig:tube_diagram}c). For both VW and TWs, the total energy density (bottom graph in Fig. \ref{fig:tube_diagram}b) closely follows the trend of the exchange energy density $E_{ex}$, implying that the impact of curvature can be primarily understood by examining its effect on $E_{ex}$.

\begin{figure}
    \centering
    \includegraphics[width=\linewidth]{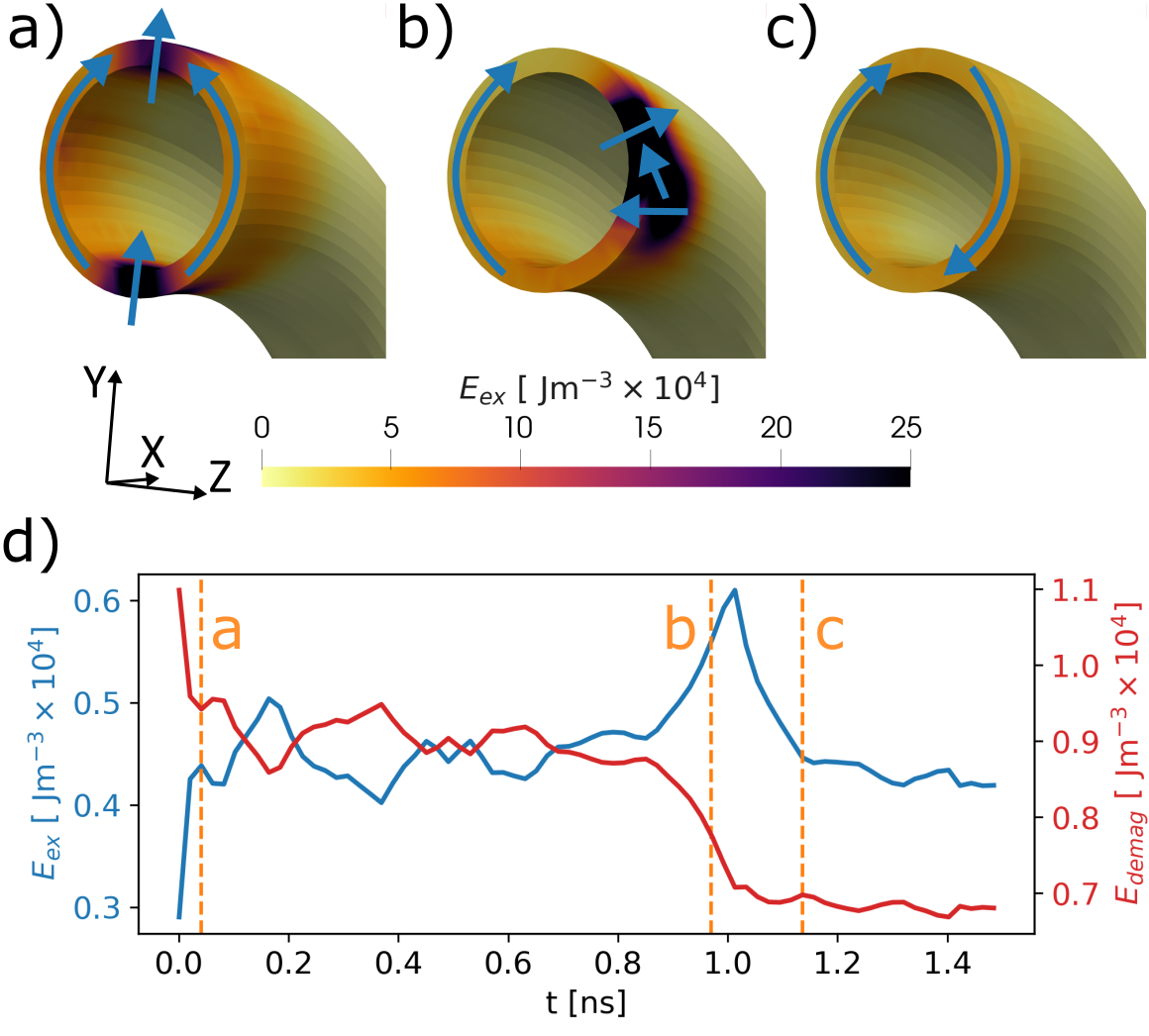}
    \caption{Dynamic collapse of the TW to a VW. (a-c) Cross-section of a nanotube with $R_C=130\unit{nm}$, $d=46\unit{nm}$ (within gray region in Fig. \ref{fig:tube_diagram}a) during the collapse, colored by the local exchange energy density. Blue arrows denote magnetization direction.  (a) The two opposite magnetic charges at the top and bottom of the tube attract and (b) move towards each other. (c) The charges annihilate resulting in the VW. (d) Evolution of the exchange and demagnetizing energy during the collapse. Dashed orange lines denote the snapshots shown in (a-c)
    }
    \label{fig:dw_collapse}
\end{figure}

As well as the relative energies of domain wall types, the stability — and in particular the likelihood of an abrupt transformation between DW types to occur — is key to understanding these systems. The large area of the diagram (both red and blue regions) where both walls are metastable implies the existence of an energy barrier for the transformation between the two configurations.
Here, we observe that the curvature affects this barrier by extending the range of diameters where the TW can exist (red and blue region in Fig. \ref{fig:tube_diagram}a).

We examine the metastability in more detail by considering the dynamic transformation process where an initialized TW state collapses to the lower energy VW.
For this, we run additional simulations for a wire in the gray region ($R_C=130\unit{nm}$, $d=46\unit{nm}$) with a realistic damping $\alpha = 0.02$, and study the time evolution of the magnetic state during the DW transformation process (Fig. \ref{fig:dw_collapse}).
We observe how the two diametrically opposed magnetic charges of the TW attract each other magnetostatically  (Fig. \ref{fig:dw_collapse}a), releasing spin waves as they move (oscillations between a and b in Fig. \ref{fig:dw_collapse}d).
The topological defects represent a vortex — antivortex pair with opposite polarization, and therefore cannot smoothly annihilate with one another requiring the formation of a magnetization singularity — a Bloch point \cite{wartelleBlochpointmediatedTopologicalTransformations2019, tretiakovVorticesThinFerromagnetic2007, thomasTopologicalRepulsionDomain2012}. As a result, the transformation faces a topological barrier. Due to limitations in the simulation of magnetization singularities with micromagnetic simulations (discussed later) the exact location of the energy barrier is subject to a small uncertainty, however it is clear that the threshold exists.
However, approaching one another requires a more abrupt spin rotation along the azimuthal direction, that comes at a cost of increased $E_{ex}$ (region between b and c in Fig. \ref{fig:dw_collapse}d). For NTs in the red region of the diagram where TW is metastable, the increased $E_{ex}$ creates a barrier to the annihilation of magnetic charges. This makes the TW metastable over a wide range of tube diameters, collapsing only when the gain in magnetostatics can overcome the exchange barrier.
As discussed earlier, introducing curvature further increases the exchange energy required to create the vortex state, raising the barrier for the motion of magnetic charges and significantly extending the range of TW metastability, as observed in the diagram.

\begin{figure*}[ht]
    \centering
    \includegraphics[width=\linewidth]{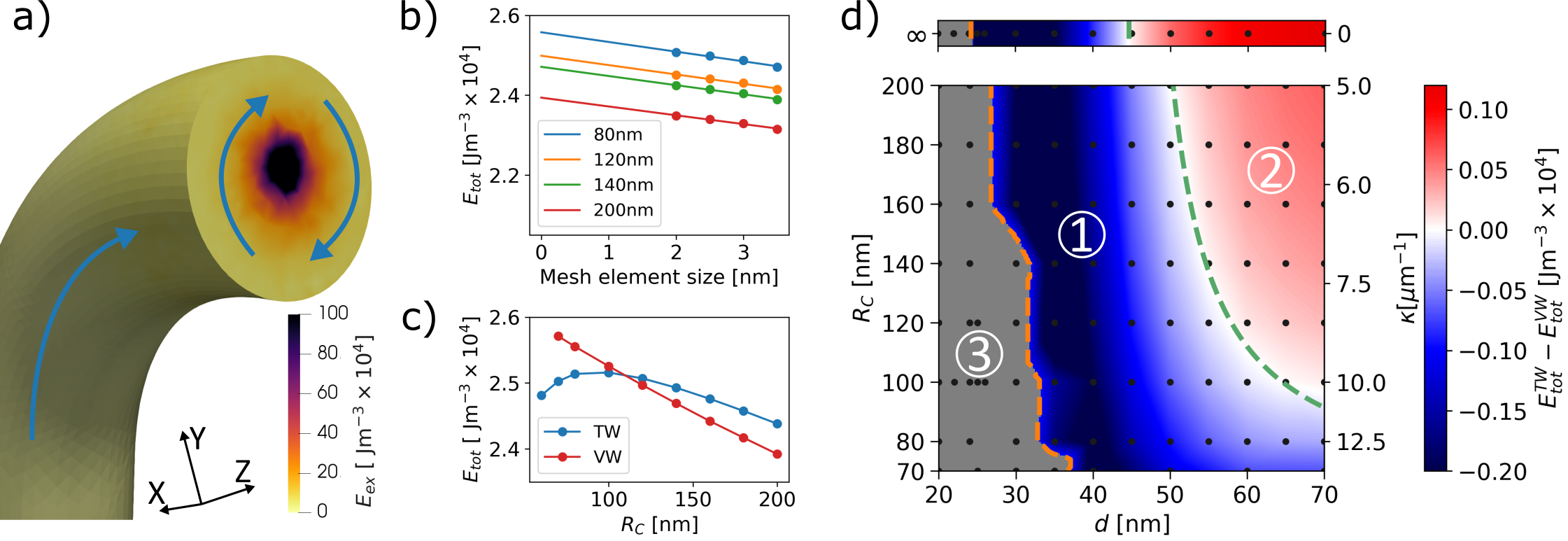}
    \caption{Domain wall curvature dependence in solid cylinders. (a) Cross-section of a curved cylinder ($R_C=100\unit{nm}, d=40\unit{nm}$)
    colored by the local exchange energy density and the magnetization direction denoted by blue arrows. The area of high exchange energy  in the middle of the wire corresponds to the region where the Bloch point singularity is positioned.
    (b) Dependence of the total energy density of the system on the mesh element size, for a wire with 60 nm in diameter and hosting a VW. The convergence to the continuum theory is acquired by extrapolating to infinite mesh density. (c) Total energy density of the VW with the Bloch point and the TW, for different radii of curvature $R_C$, in the case of the $d=60\unit{nm}$ wire. (d) Full phase diagram of the energy density difference $E^{TW}_{tot}-E^{VW}_{tot}$, showing three distinct regions. In regions 1 and 2, both domain walls can exist, with TW being favorable in the former (blue region with $E^{TW}_{tot}<E^{VW}_{tot}$) and VW in the latter (red region with $E^{VW}_{tot}<E^{TW}_{tot}$). In the gray region 3, the VW is unstable and collapses to the lower energy TW state.}
    \label{fig:bp_energies}
\end{figure*}

After studying the effect of curvature on NTs, we perform a similar investigation in solid NWs.
Even though they share many of the same features, the solid NWs are topologically different from NTs. The effect of the topology is most apparent in the case of the VW which in NWs forms a Bloch point, a micromagnetic singularity at the center of the wire (see Supporting Information S3)\cite{dacolObservationBlochpointDomain2014}.

A fundamental assumption of the micromagnetic theory is the homogeneous saturation magnetization.
At the vicinity of the BP, this assumption is broken as the magnetization is strongly inhomogeneous at the atomistic scale (Fig. \ref{fig:bp_energies}a). It has been shown that micromagnetic simulations can nevertheless describe the structure and many dynamic features involving BPs, matching well with experiments \cite{thiavilleMicromagneticStudyBlochpointmediated2003, dacolObservationBlochpointDomain2014}. However, the micromagnetic exchange energy density diverges at the singularity, requiring a careful analysis to prevent discretization-induced artifacts \cite{andreasNumericalMicromagnetismStrong2014}.


The full treatment of the problem necessitates the use of an atomistic model in the vicinity of the BP \cite{hertel22MultiscaleSimulation2015, hertelUltrafastDomainWall2016}, or an adaptation of the continuum theory to allow changes in the saturation magnetization \cite{galkinaPhenomenologicalTheoryBloch1993,lebeckiFerromagneticVortexCore2014,eliasMagnetizationStructureBloch2011,garaninFokkerPlanckLandauLifshitzBlochEquations1997}.
Another common way to address the issue of singularities is to run the simulations with different mesh element sizes and extrapolate the computed values to an infinite mesh density \cite{andreasNumericalMicromagnetismStrong2014,gligaEnergyThresholdsMagnetic2011, thiavilleMicromagneticStudyBlochpointmediated2003, hertelFiniteElementCalculations2002}. The energy obtained in this way converges to the continuum theory, removing the dependence on the underlying mesh.
We employ the latter method (Fig. \ref{fig:bp_energies}b) to demonstrate a similar curvature dependence of DWs in solid cylinders as in tubes.
It is important to keep in mind that the non-physical divergence of the exchange energy density is still present in the continuum theory, which can shift the overall energy of the VW containing the BP. However, as the singularity is embedded in the volume, the shift only depends on the material parameters and not the external geometry, allowing for a valid geometrical comparison as performed here.

To understand the energetics of both types of walls in NWs, we focus first our discussion on the particular case of a nanowire with $d = 60 \unit{nm}$ (Fig.  \ref{fig:bp_energies}b-c). We plot $E_{tot}$ as a function of curvature (Fig. 4c), observing a similar behavior to the one found for NTs (bottom plot in Fig. \ref{fig:tube_diagram}b).
As the radius of curvature decreases, the energy cost of the VW steadily increases, while the TW energy sharply drops for very curved wires. This is unsurprising, as the same interpretation as for the NTs, with curvature-induced DMI and anisotropy promoting the TW state, is still valid on the outer sections of the cylinder.

Having discussed a particular case that demonstrates the similarity between NTs and NWs, we now simulate the full phase diagram of TWs and VWs in curved NWs (Fig. \ref{fig:bp_energies}d).
We initialize the wires and calculate the DW energy following the same procedure as in NTs, consisting of transforming the mesh and magnetization from the straight geometry. To be able to reliably initialize both types of DW, we use $d_0=50 \unit{nm}$ for the initial diameter. As explained before, for the VWs with BPs, we further run the simulations with multiple mesh sizes, interpolating the energy to infinite mesh density (Fig. \ref{fig:bp_energies}b).

When compared to the case of tubes, we observe in NWs an increase in the energy of the VW over the TW, leading to the TW being the preferred state over a larger range of diameters. For straight wires, the TW is a favored state up to 45 nm, a significant increase from 24 nm in the tubes (green line in Fig. \ref{fig:bp_energies}d). Additionally, for curvature radii below $R_C=90\unit{nm}$, the TW is the ground state for the full range of investigated diameters.
Furthermore, while for tubes the TW was unstable at high diameters with the spontaneous collapse to the VW being suppressed by the curvature, the contrary happens in NWs. The presence of a BP plays thus a key role as well in the range of metastability of the VW in wires. Here, the VW collapses to a TW for diameters below $d=24 \unit{nm}$ in straight wires, further increased by curvature to $d=35\unit{nm}$ at $R_C = 70 \unit{nm}$ (orange line in Fig. \ref{fig:bp_energies}d).

The transformation of VWs into TWs in NWs due to curvature, and the inhibition of the inverse transformation in NTs, are particularly relevant when considering these systems as domain wall conduits for sensing and memory devices \cite{parkinMagneticDomainWallRacetrack2008,borieGeometricalDependenceDomainWall2017}. For instance, in the case of NWs described last, while the VW containing a BP can be metastable in a wire with $d=30\unit{nm}$, it will be energetically favorable to spontaneously collapse to the TW state when located within a region with radius of curvature $R_C \leq 140\unit{nm}$. The resulting TW is now in its ground state and remains metastable even at significantly higher diameters, making it hard to induce an inverse transformation. Additionally, our results indicate how curvature can play a key role for the control of the stability and pinning of DWs, tuning their stochastic behavior in proposed neuromorphic devices \cite{dawidekDynamicallyDrivenEmergenceNanomagnetic}.

To summarize, we demonstrate the effect of curvature on the energy and stability of magnetic domain walls in cylindrical and tubular nanowires. We map the phase diagram of the relative energies of transverse and vortex domain walls for varying diameters and curvatures. We identify the presence of a region of metastability where both walls can exist, as well as the boundary where the transition between the ground DW states takes place. In both geometries, introducing curvature increases the energy of the vortex wall while decreasing the energy of the transverse wall. This promotes the stability of the transverse wall, making it a favorable state for a larger range of diameters, inhibiting its collapse in NTs, and making VW collapse more likely in NWs. This behavior is understood to be mainly due to the curvature-induced effective DMI driven by the exchange energy.
These results are key for the fundamental understanding of domain walls in curved geometries, and for device applications.

\begin{acknowledgments}
    This work was supported by the EPSRC Cambridge NanoDTC EP/L015978/1, and the Winton Program for the Physics of Sustainability. L. Skoric acknowledges support from St Johns College of the University of Cambridge.
    C. Donnelly  was supported by the Leverhulme Trust (ECF-2018-016), the Isaac Newton Trust (18-08) and the L'Or\'eal-UNESCO UK and Ireland Fellowship For Women In Science.
    A. Hierro-Rodriguez acknowledges support from Spanish AEI under project reference PID2019–104604RB/AEI/10.13039/501100011033.
    The authors acknowledge the University of Vienna research platform MMM Mathematics - Magnetism - Materials, and the FWF project I 4917.
\end{acknowledgments}

\bibliography{curved_nanowires}

\end{document}